# In-situ observation of field-induced nano-protrusion growth on a carbon-coated tungsten nanotip

Guodong Meng[1,*], Yimeng Li[1], Roni Aleksi Koitermaa[2,3], Veronika Zadin[2], Yonghong Cheng[1], Andreas Kyritsakis[2,3,†]

[1]*State Key Laboratory of Electrical Insulation and Power Equipment, Xi'an Jiaotong University, Xi'an, China*
[2]*Institute of Technology, University of Tartu, Nooruse 1, 51010 Tartu, Estonia*
[3]*Department of physics, University of Helsinki, PO box 43, FI-00014 Helsinki, Finland*

Nano-protrusion (NP) on metal surface and its inevitable contamination layer under high electric field is often considered as the primary precursor that leads to vacuum breakdown, which plays an extremely detrimental effect for high energy physics equipment and many other devices. Yet, the NP growth has never been experimentally observed. Here, we conduct field emission (FE) measurements along with in-situ Transmission Electron Microscopy (TEM) imaging of an amorphous-carbon (a-C) coated tungsten nanotip at various nanoscale vacuum gap distances. We find that under certain conditions, the FE current-voltage (*I-V*) curves switch abruptly into an enhanced-current state, implying the growth of an NP. We then run field emission simulations, demonstrating that the temporary enhanced-current *I-V* is perfectly consistent with the hypothesis that a NP has grown at the apex of the tip. This hypothesis is also confirmed by the repeatable in-situ observation of such a nano-protrusion and its continued growth during successive FE measurements in TEM. We tentatively attribute this phenomenon to field-induced biased diffusion of surface a-C atoms, after performing a finite element analysis that excludes the alternative possibility of field-induced plastic deformation.

The application of intense electric fields on a metal surface can cause vacuum breakdown (VBD), also known as vacuum arc [1]. This severely hinders the performance of several devices, including vacuum interrupters [2], X-ray sources [3], fusion reactors [4] and particle accelerators such as the existing Large Hadron Collider [5] and the next-generation Compact Linear Collider (CLIC) [6, 7]. These issues cannot be mitigated yet, due to insufficient scientific understanding on the mechanisms that cause VBD. Generally, it is widely accepted [1, 8-11] that the initial vapor and ion population that leads to plasma onset is produced by the atom evaporation due to extreme heating, induced by a localized field electron emission spot that enters a thermal runaway process [8, 9, 11-14]. However, for such a process to occur, an extremely high geometric field enhancement (of the order of several hundreds) by localized sharp protrusions on the metal surface needs to be assumed [15], which have not been observed experimentally especially for the metal surface after prior conditioning, so it can be suggested that these sharp protrusions on metal surfaces are created spontaneously under high electric field [16]. Experience from industrial production suggests that the diffusion of the adsorbates or contaminants such as carbon compounds etc. on the metal surface may play a dominant role in the extremely high geometric field enhancement (build-up of nano-protrusions (NP)) and subsequent electrical breakdown under high electric field despite in the vacuum environment, which can cause the cathode craters [17-21]. Yet, the exact mechanisms that lead to this phenomenon have not yet been understood, and its relevance to VBD condition has not been demonstrated.

In terms of the exact mechanisms of the appearance of field enhancing protrusions, several processes have been proposed, including plastic deformations due to high electric field-induced stress [22-24], field-induced biased surface diffusion [25]. Yet, none of them has been proved experimentally or described accurately in theory, due to the high complexity of the field-induced modifications of the interatomic interactions. Several experiments have demonstrated that the diffusion of surface adsorbates or contaminants under high electric field can lead to the build-up of nano-protrusions (NPs) under high electric field during field emission (FE) [19, 26, 27]. This process, apart from the possibility to lead to vacuum arc, can be used to achieve extremely high-performance electron emitters [28, 29]. Yet, the exact mechanisms that lead to this phenomenon have not yet been understood, raising the need for in-situ studies of NP growth under high electric field.

In this letter, we study in-situ the morphological evolution of a field emitting tungsten (W) nanotip coated with an amorphous carbon (a-C) layer during the field emission process. We measure the evolution of its field emission characteristics in-situ a transmission electron microscope (TEM) and record the growth of an a-C nano-protrusion, which is attributed to field-induced biased surface diffusion. This is the first in-situ observation of field-induced NP growth on an a-C layer coated metal surface, demonstrating the possibility of such a process leading to protrusion growth and revealing the origin of enhanced field emission before vacuum arcs.

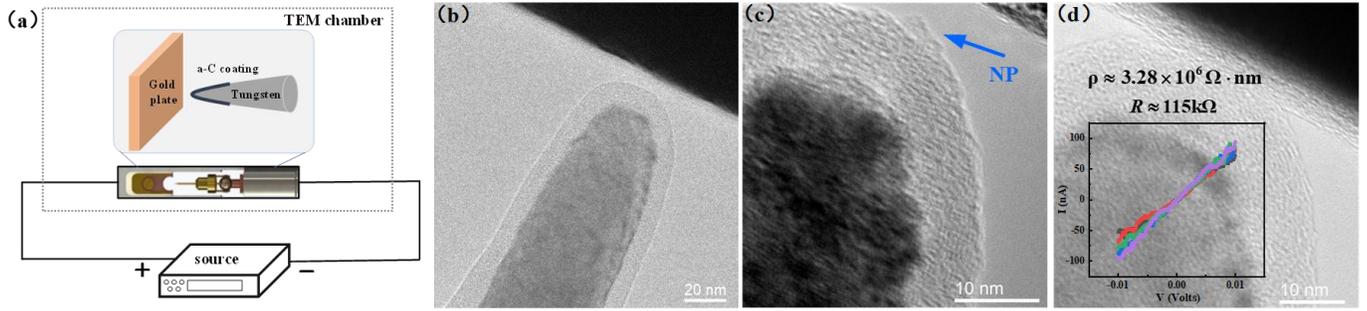

FIG. 1. (a) Schematic diagram of the in-situ morphology characterization and field emission measurement system. (b) TEM image of the amorphous carbon-coated tungsten nanotip and the gold plate anode. (c) TEM images of a-C coated tungsten nanotip after the FE measurements in $d_3$ gap and the corresponding nano-protrusion growth. (d) TEM image of the nanotip and anode in contact. Inset in (d): I-V curves during short circuit with the corresponding resistivity of the coating being ~3.28×10$^6$ Ω·nm.

The in-situ field emission experiments and corresponding electrode morphology recordings are carried out inside a JEOL-2010F TEM, where the pressure is on the order of 10$^{-5}$ Pa, and the electrode gap is adjusted by an in-situ electrically biased TEM holder (ZepTools Technology) as shown in the schematic of Fig. 1(a). The W tip was electrochemically etched down to ~20 nm radius in a NaOH/KOH solution [30]. Then an a-C thin layer was deposited onto the W nanotip by electron-beam-induced deposition. Fig. 1(b) shows the TEM image of an as-prepared a-C coated W nanotip with about 29 nm radius. Fig. S1 in the Supplemental Material presents the results of energy dispersive spectroscopy (EDS) element analysis and electron diffraction (SAED) analysis for different selected areas. The EDS result confirms that the tip core is W, and the shell is carbon. A single crystal grain is observed at the W apex area based on the SAED result. Fig. 1(d) shows the TEM image and I-V curves when the a-C coated tungsten nanotip is brought to contact with the gold-plate counter-electrode. The short circuit current shows a linear relationship as a function of the applied voltage, and the volumetric resistivity could be derived to be about 3.28×10$^6$ Ω·nm, exhibiting graphite-like properties [31, 32].

We took field emission I-V measurements for various gap distances $d$. Unfortunately, this particular setup does not allow for an accurate estimation of $d$ from the TEM images, due to the high sensitivity of this measurement to even minute tilting of the anode plate (see the Fig. S2 in the Supplemental Material). In several cases, we observed that the field emission process induced significant changes in the morphology and the emission behavior of the tip. Although the exact dynamic process of these morphological changes is extremely difficult to observe in real time by TEM, a careful analysis of the field emission behavior of the tip can give us significant insights.

Fig. 2 presents four consecutive sets of I-V measurements, with various gap distances. We see that for $d_1$ and $d_2$, the I-V curves exhibit a metastable behavior, evolving through four distinct states. In State #1, the tip emits a low current ($I_{low}$), following an approximately stable Fowler-Nordheim (FN) type curve. In the following State #2, the I-V starts with a similar characteristic at low voltages, but when the current reaches the 10 nA scale, it abruptly jumps to a significantly higher current ($I_{high}>4I_{low}$). Then, in State #3, the I-V continues following the high current behavior, in a line consistent with the formation of an NP that slightly increases the local electric field at the tip. Finally, in State #4, the emission returns to $I_{low}$, following a similar characteristic as in State #1. Here we use State #4 instead of State #1 to illustrate the temporal evolution, indicating that the FE-enhancing features that cause the high-current state (probably an NP) can appear and disappear, which indicates a metastable process.

For the $d_3$ gap, the I-V curves have only two distinct states. One is similar to State #2, i.e., the current starts at a low level and then is enhanced up to four times at higher voltages, but in this case the I-V curve has significant fluctuations at higher voltages rather than jumping to a stable high-current state, indicating an extremely unstable a-C layer. The other is similar to the stable enhanced-current State #3. Then, with progressive FE measurements at a reduced nanogap $d_4$, an NP was observed to be formed firmly near the tip apex, and only the stable high-current State #3 was obtained afterwards.

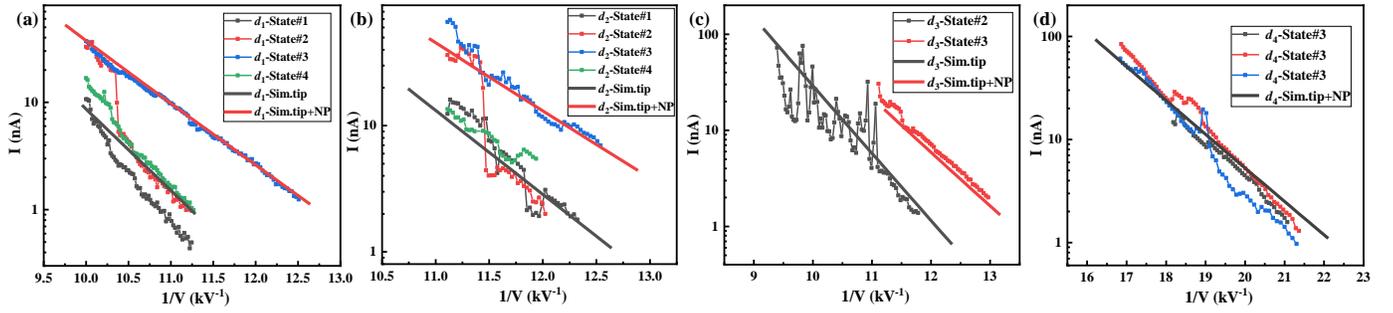

FIG. 2. Measured field emission *I-V* curves (dotted lines) from the a-C coated tungsten nanotip under different nanogaps. The corresponding simulated curves are given in solid lines, with the fitted gap distances found to be $d_1$=50 nm, $d_2$=37 nm, $d_3$=41.5 nm, $d_4$=17 nm. The field emission current ($d_{1,2}$) exhibits four states: initial stable low current State #1, low-high current transitional State #2, stable high current State #3, and final stable low current State #4. In the $d_3$ case, the field emission current shows two states: low-high current transitional State #2 with an intensively fluctuating *I-V*, and stable high current State #3. In the $d_4$ case, only the stable high-current State #3 is observed, together with a permanent NP outgrowth visible in TEM.

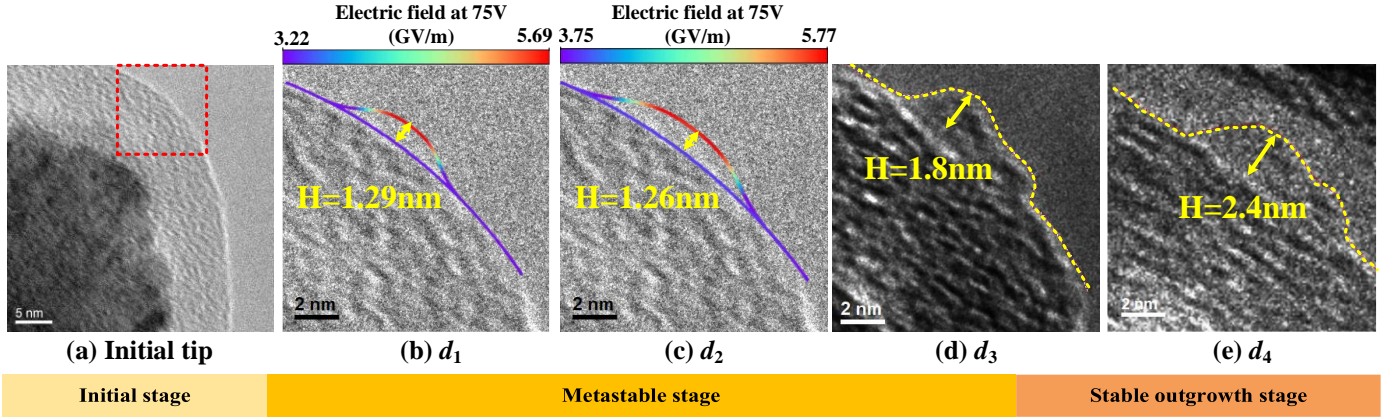

FIG. 3. Schematic representation of the tip evolution. (a) the initial nanotip before FE measurements. (b-c) the simulated NP (colored lines) for $d_{1,2}$. The color coding gives the calculated surface electric field distribution, with the NP exhibiting a significant local field enhancement. (d-e) the stable NP outgrowth visible in TEM after the FE experiments in $d_3$ and $d_4$.

To test our NP growth hypothesis as an explanation for the transitional FE behavior of Fig. 2, we simulated the FE process fully, using our multi-physics software FEMOCS [9, 12, 13], which is coupled with the general field emission calculation software GETELEC [33]. We simulated a static tip with a hemisphere-on-cone geometry, which matches the TEM images of the emitter. The tip radius was set to *R*=29 nm and the shank cone full-aperture angle to 2°. The electric field is calculated around the tip by solving the Laplace equation using the Finite Element Method (FEM). Then the electrostatic potential near the tip surface is passed to GETELEC, which calculates the FE current density distribution around the tip and the corresponding total emitted current (see Refs. [9, 33] for details), taking into account the local emitter curvature [34]. The work function of the emitter was set to the one of graphite, $\phi$ = 4.62 eV [35]. The total current is multiplied by a fitted pre-exponential correction factor of 0.025, which accounts for various uncertainties in the field emission theory (e.g., barrier shape, tunneling probability, real band structure, etc.) [36].

For the measurements at $d_{1,2,3}$, we simulated two geometries. One with the regular smooth tip and one with a small NP at the apex, as illustrated in Fig. 3(b-c). For the measurements at $d_4$, only the geometry with NP was simulated. The value of the anode-cathode distance was fitted for the smooth tip case to the experimental low-current curve, i.e., State #1, being found $d_1$=50 nm, $d_2$=37 nm, $d_3$=41.5 nm, $d_4$=17 nm. The resulting simulated *I-V* for are plotted together with the experimental ones in Fig. 2. We see that the curves without the NP are in good agreement with the low-current curves (States #1,4), while the one with the NP matches very well with the enhanced-current State #3. The height of the NP is chosen to fit the third *I-V* curve for State #3, being H=1.29 nm for $d_1$, and H=1.26 nm for $d_2$. The simulated NP shape for these cases is shown in Fig. 3(b-c). An important characteristic of the simulated *I-V*, that agrees with the experimental measurements, is that the one with the NP corresponds to a slightly higher field enhancement factor and a significantly lower effective emission area, obtained upon fitting the simulated curves to the standard Muprhy-Good equation using GETELEC [33] and its on-line interface [37]. The obtained values are shown in Table I. Here the enhancement factor is defined as $\beta$=*Fd/V*, with *F* being the electric field entering the Murphy-Good equation. Furthermore, based on the simulated field enhancement factors, the maximum electric fields during the *I-V* measurements can be obtained, as the Table I shows.

TABLE I. Extracted effective enhancement factors for the tip without NP ($\beta_{tip}$) and tip with NP ($\beta_{NP}$), and emission areas for the tip without NP ($A_{tip}$) and tip with NP ($A_{NP}$) from the simulated I-V curves of Fig. 2.

| Gap distance (nm) | $\beta_{tip}$ | $\beta_{NP}$ | $A_{tip}$ (nm$^2$) | $A_{NP}$ (nm$^2$) |
| --- | --- | --- | --- | --- |
| $d_1$(50nm) | 1.975 | 2.6 | 3743 | 237 |
| $d_2$(37nm) | 1.67 | 2.1 | 2980 | 315 |
| $d_3$(41.5nm) | 1.77 | 2.26 | 3305 | 266 |
| $d_4$(17nm) | - | 1.58 | - | 159 |

The height of the NP first observed experimentally is about 1.8 nm (Fig. 3(d)). With further field emission at $d_4$, the NP height increased slightly to 2.4 nm (Fig. 3(e)). Our results indicate that the simulated NPs for $d_{1,2}$ are metastable, with a height of about 1.3 nm, thus they cannot be observed in TEM. The exact physical process of NP growth is not clear yet. However, we can assume its basic characteristics based on the observed FE dynamics. The I-V curves in State #2 exhibit a quick transition upon the application of high field that draws high FE current, while in other circumstances they transit to a relatively stable condition. This implies that the physical process that causes these transitions possibly involves the effects of high electric field on the surface (e.g., Maxwell stress and modification of the interatomic interactions [39]). Additionally, an important role can be attributed to the local heating due to the field emission current, caused by the Nottingham and Joule effects. Yet, the temperatures cannot reach near the melting point of a-C because this would probably induce an extremely rapid and violent thermal runaway process [9]. There are two known mechanisms that involve the above processes. One is plastic deformation due to the Maxwell stress and the second is surface diffusion, biased by the modification of the interatomic interactions due to the high electric field [25, 40]. One could also consider deposition of adsorbents from the vacuum or from electron-induced degassing of the anode [20], but this mechanism was ruled out by additional experiments on clean W tips under the same setup that remained clean after a long stable emission (see the Fig. S7 in the Supplemental Material).

To test the validity of the plastic deformation hypothesis, we conducted FEM simulations utilizing our previously developed electrostatic-elastoplastic model [16, 41] based on the constitutive modelling approach for continuum mechanics [42]. Continuum methods rely on the homogenization of the material behavior over large volumes. Although this is not an accurate approach if only a few atoms are involved, reliable order of magnitude estimations can be made [43]. Here we consider the elastoplastic deformation of the a-C tip coating under the electric field induced Maxwell stress. To parameterize the elastoplastic properties of a-C, we fitted them against the nano-indentation measurements of Ref. [44] and the simulation results of Ref. [40]. The aim of our study was to evaluate if the Maxwell stress can cause yielding and plastic deformation consistent with the observed surface modifications.

The results of our model are summarized in Fig. 4. We first calculated the electric field distribution, which is shown in the "hot" color scale around the tip. The anode-cathode distance was set to $d_1$=50 nm and the applied voltage was ramped from 0 to 170V to cover the theoretical range of 8 GV/m maximum local field on the tip, which is an upper limit for the field magnitude of our FE simulations. We then calculated the distribution of the total strain and stress in the coating layer at 100V applied voltage, which exceeds the maximum of the experiments and corresponds to apex field of 4.71 GV/m. The stress distribution is depicted with the "rainbow" color scale, while the "jet" one gives the percentage of the plastic component of the total strain. We see that the contribution of the plastic deformation on the total strain is less than 0.001, with the stress being significant only at the inner part of the a-C layer, where the high-stiffness high-curvature W core acts as a stress concentrator. The apex area of tip becoming plastic (non-zero plastic strain) around $V$~150V or $F_{max}$~7GV/m, which far exceeds the voltage applied in our experiments. This means that the a-C coating is expected to behave almost entirely elastically within the range of stresses exerted to it by the electric fields involved in FE experiments. We repeated the same calculation for various distances (see the Fig. S8 in the Supplemental Material), yielding similar results. The above results effectively exclude pure plastic deformation as the cause of at least the initiation of the NP growth we observed. It would be very hard to consider a plausible scenario by which the field-induced stresses within the FE range can cause significant plastic deformation to the a-C coating.

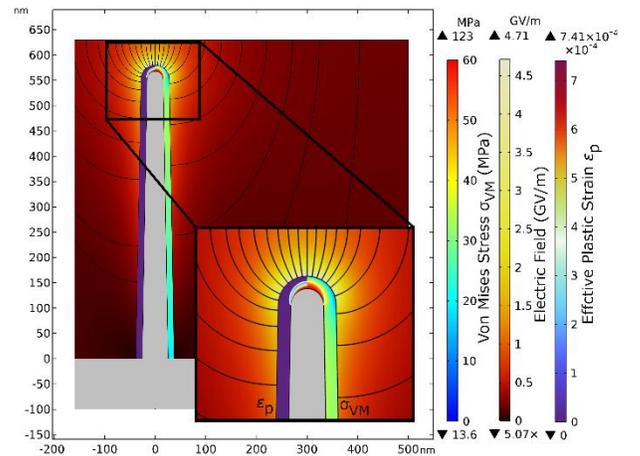

FIG. 4. Results of the electrostatic-elastoplastic model for the a-C coating under Maxwell stress.

This implies that the most plausible mechanism behind the observed surface modifications is field-induced biased surface diffusion. The high local electric field significantly modifies the migration barriers of surface atoms for any metal [39, 40, 46-48], causing the diffusion of surface atoms to be biased towards higher field values. This effect has been shown to induce significant modification on metal surfaces, although typically very high fields (beyond the FE range used here) and temperatures (close to melting) are involved [25, 49, 50]. However, it is plausible to assume that the atoms on the surface of the a-C coating are more loosely bound (especially π-bonds) than those of a metal, leading to an increased mobility of the surface atoms that allows the biased surface diffusion to become significant at moderate temperatures.

The surface diffusion mechanism is also compatible with the observed instability of the NP growth. The FE dynamics of Fig. 2 imply that the NP grows under certain conditions when a significant FE current is drawn, but then flattens back to its original state after the FE experiment ceases, which is compatible with surface diffusion dynamics. When no field-induced bias is present on the surface, diffusion typically promotes the flattening of any asperity, as it seeks to minimize the surface energy and maximize the volume to surface ratio. However, upon application of high field and possibly moderate local heating caused by FE, the biased diffusion can induce NP growth due to the diffusion bias towards a positive field gradient [39, 40, 46].

This diffusion hypothesis cannot yet be confirmed by computational models. Moreover, it remains an open question what is the mechanism that stabilizes the NP after the measurements at $d_3$, making it visible on the TEM image. We can hypothesize that under specific circumstances the atoms on the NP might take metastable configuration, needing to overcome a relatively high barrier to flatten out. Yet, we currently lack the tools to test this. Although previous attempts for kinetic Monte Carlo (KMC) simulations that include the field effects have been made [25], their application to a non-metal surface such as a-C is very complex. Furthermore, even for simple metals, the exact modification of the interatomic interactions by the field is highly dependent on the local atomic environment, which further complicates their consideration in KMC simulations [51]. It is therefore essential and timely to develop such models, test the biased diffusion hypothesis and understand its dynamics quantitatively.

Furthermore, in addition to the observation of NP under $I$-$V$ conditions (ramping voltage measurement) in this paper, we have also observed the dynamic growth of NP during a constant voltage measurement in other experiments, which was accompanied by the enhanced field emission current and induced a following electrical breakdown (see the Figs. S3 to S6 in the Supplemental Material).

In conclusion, we demonstrated a nano-protrusion growth on the amorphous carbon (a-C) coating layer of a tungsten nanotip during field emission, by in-situ field emission experiments and accompanying field emission simulations. We attribute the nano-protrusion growth to field-induced biased surface diffusion, after excluding the possibility of field-induced plastic deformation by FEM simulations. This letter provides the experimental confirmation of amorphous-carbon nano-protrusion growth on an electrode surface exposed to high electric field, which is a plausible explanatory mechanism of the appearance of field enhancing features necessary to initiate electrical breakdown in vacuum.

ACKNOWLEDGMENTS

The authors are grateful to The Center for Advancing Materials Performance from the Nanoscale (CAMP-Nano) in Xi'an Jiaotong University for the in-situ TEM measurement, and the ZepTools Technology Company for the TEM holder. This work was supported by the National Natural Science Foundation of China (51977169), European Union's Horizon 2020 research and innovation programme ERA Chair "MATTER" (grant No. 856705), and by the Estonian Research Council's project RVTT3 – "CERN Science Consortium of Estonia".

*gdmengxjtu@xjtu.edu.cn
†andreas.kyritsakis@ut.ee

# In-situ observation of field-induced nano-protrusion growth on a carbon-coated tungsten nanotip
## Supplemental Material

Guodong Meng[1,*], Yimeng Li[1], Roni Aleksi Koitermaa[2,3], Veronika Zadin[2], Yonghong Cheng[1], Andreas Kyritsakis[2,3,†]

[1]*State Key Laboratory of Electrical Insulation and Power Equipment, Xi'an Jiaotong University, Xi'an, China*
[2]*Institute of Technology, University of Tartu, Nooruse 1, 51010 Tartu, Estonia*
[3]*Department of physics, University of Helsinki, PO box 43, FI-00014 Helsinki, Finland*

*gdmengxjtu@xjtu.edu.cn
†andreas.kyritsakis@ut.ee

# List of supporting video

**Video S1.** The NP growth during a constant voltage measurement (V=125V).

## The characterization of the carbon-coated tungsten nanotip

The Fig. S1 presents the results of energy dispersive spectroscopy (EDS) element analysis and electron diffraction (SAED) analysis for different selected areas of this nanotip. The EDS result confirms that the tip core is W, and the shell is carbon. A single crystal grain is observed at the W apex area based on the SAED result.

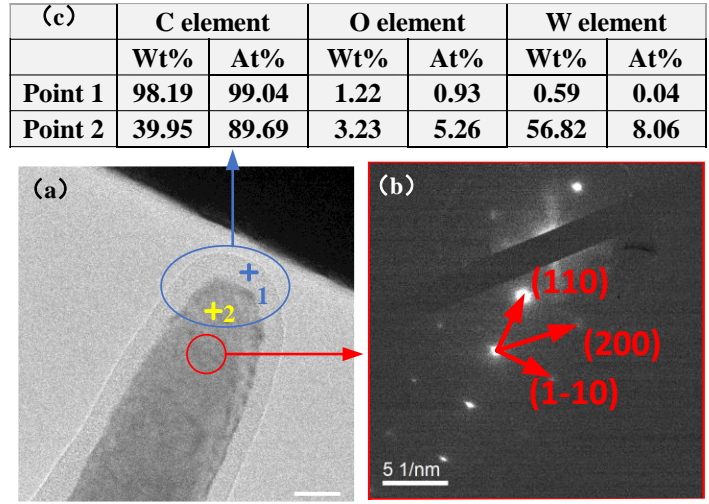

FIG. S1. (a) TEM image of the amorphous carbon coated tungsten nanotip and the gold plate anode; (b) The selected area electron diffraction (SAED) pattern of the tip; (c) The results of EDS elemental analysis on two different points.

## The underestimation of gap distance

As the Fig. S2 shows, the anode plane is about 100 μm in width and 500 μm in length. Since the apparent anode-cathode distance in the TEM image is given by the projection of the anode plane to the e-beam, it is highly dependent on the tilt angle between the anode plane and the e-beam. The real distance is $d_r = d_a + h \sin\theta$, where $d_a$ is the apparent distance in the TEM image and $\theta$ is the tilt angle. If $h$ is the half thickness of the anode plane, an uncertainty as low as 0.05 degrees in $\theta$, translates in an uncertainty of about 44nm in the estimation of distance.

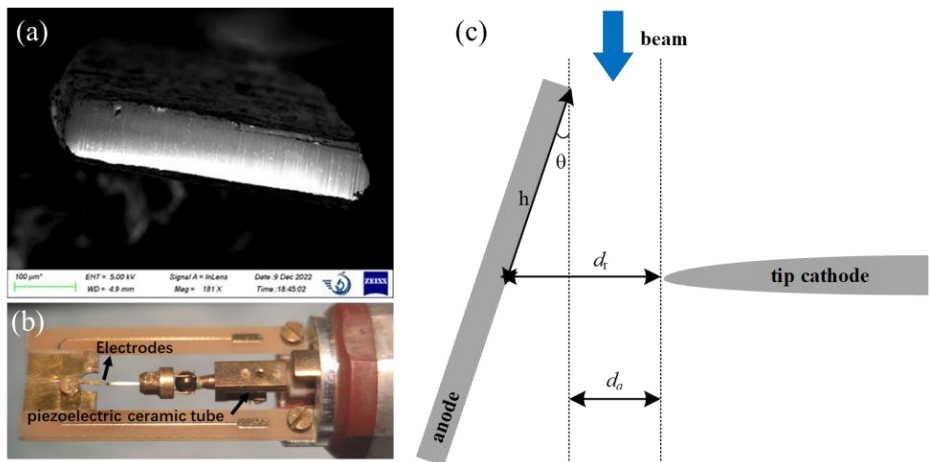

Figure S2. (a) The SEM image of anode plane; (b) the TEM holder and the electrodes; (c) the schematic diagram of the gap distance.

## The dynamic growth of NP during the constant voltage measurement

We have conducted more experiments on other carbon-coated tungsten nanotips, and reproduced the similar behavior of NP growth under FE conditions.

The morphology of this carbon-coated tungsten nanotip along with the corresponding *I-V* curves is shown in Figure S3. During our initial measurements with *ramping voltage*, the a-C layer exhibited a very stable behavior, with the field emission *I-V* curves being stable at various nanogaps. The morphology of this nanotip remained basically unchanged after these field emission *I-V* measurements.

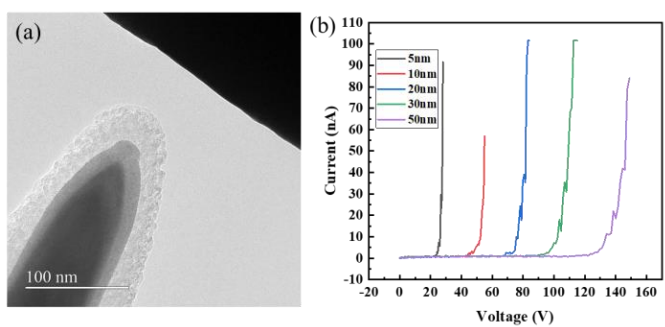

Figure S3. (a) The TEM image of the carbon-coated tungsten nanotip; (b) the field emission *I-V* curves at different nanogaps.

Next, we ran a *constant voltage measurement* (V=125V, d=50nm) for a longer time (~5 minutes) and observed the whole dynamic evolution process of the a-C surface (Video S1-d50nm-NP Growth). Figure S4 shows the typical frames and corresponding field emission currents, and it can be seen that surface becomes gradually rougher, with a clear nano-protrusion growth at the apex, accompanied by an increasing evolution of the field emission characteristics, where the field emission current increases from 2 nA to 101 nA. In addition, the *I-V* curves before and after NP growth were also measured and compared (see Figure S5), indicating that the carbon-coated nanotip underwent an increase of its field enhancement factor during the FE process.

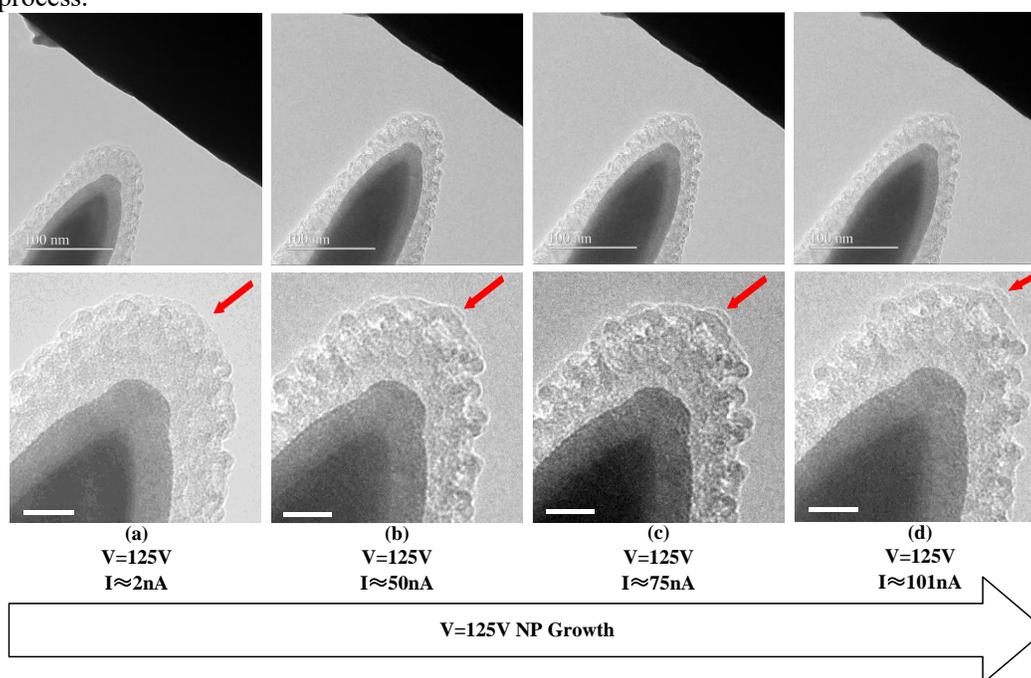

Figure S4. Characteristic frames and corresponding field emission currents during the *constant voltage measurement*

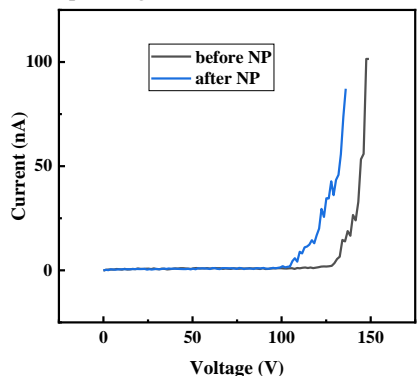

Figure S5. The IV curves before and after NP growth

Finally, upon continuing the experiment, an electrical breakdown occurred. The electrodes' morphology after the electrical breakdown are shown in Figure S6. It can be seen that as the NP grows, the growing NP under high electric field eventually induces the electrical breakdown, and the cathode of carbon-coated tungsten nanotip melts into mushroom heads, and the anode surface becomes rough.

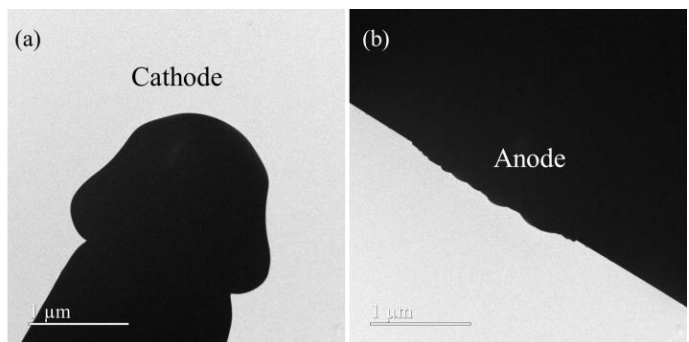

Figure S6. The TEM images of the electrodes' morphology after electrical breakdown.

Therefore, the key components of the dynamic NP growth behavior were obtained in multiple samples, and the accompanied field emission enhancement were also observed. In addition, the growing NP that could induce the electrical breakdown was verified.

## The exclusion of the deposition mechanism

The experimental setup is very similar to the one used at the *constant voltage measurement*, with the difference that we use a clean crystalline W tip. Although there are the adsorbates on the counter electrode or gas in the chamber (especially hydrocarbons), the results show that with the continuous field emission, although the morphology of pure tungsten nanoelectrode has evolved, and the internal lattice structure remained regular, no pollutants such as carbon layer appeared on its surface, as shown in Figure S7 below. This indicates that the adsorbed gas on the anode plate or the residual gas in the vacuum cavity cannot form the carbon nano-protrusion directly, at least not by themselves, and a significant component of surface self-diffusion on the a-C layer should be invoked to explain our experiments.

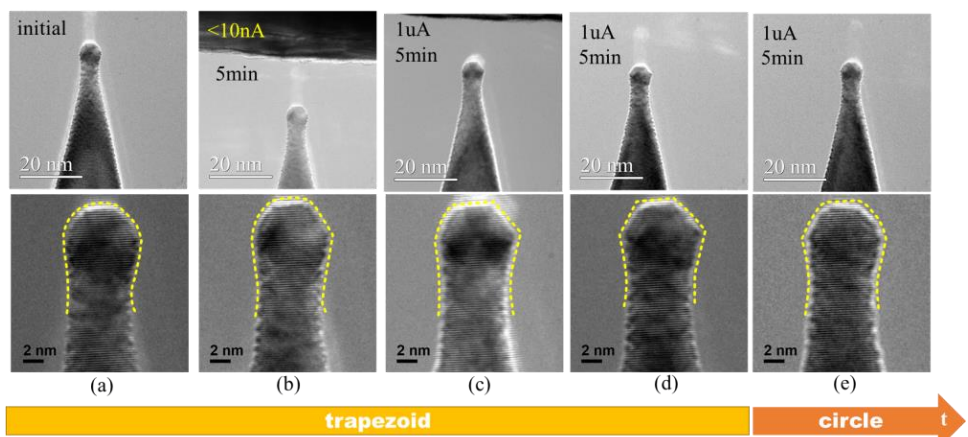

Figure S7. The morphology changes of the pure tungsten nanotip during the constant voltage measurement.

## Plastic deformation calculations in various distances

Repeating the simulation of plastic deformation for various gap distances, and the results are shown in Figure S8. The images have the same format as Fig. 4 in the main text. In all cases the plastic deformation strain does not exceed 0.00075. The white line shows the boundary of the region where the plastic deformation is zero.

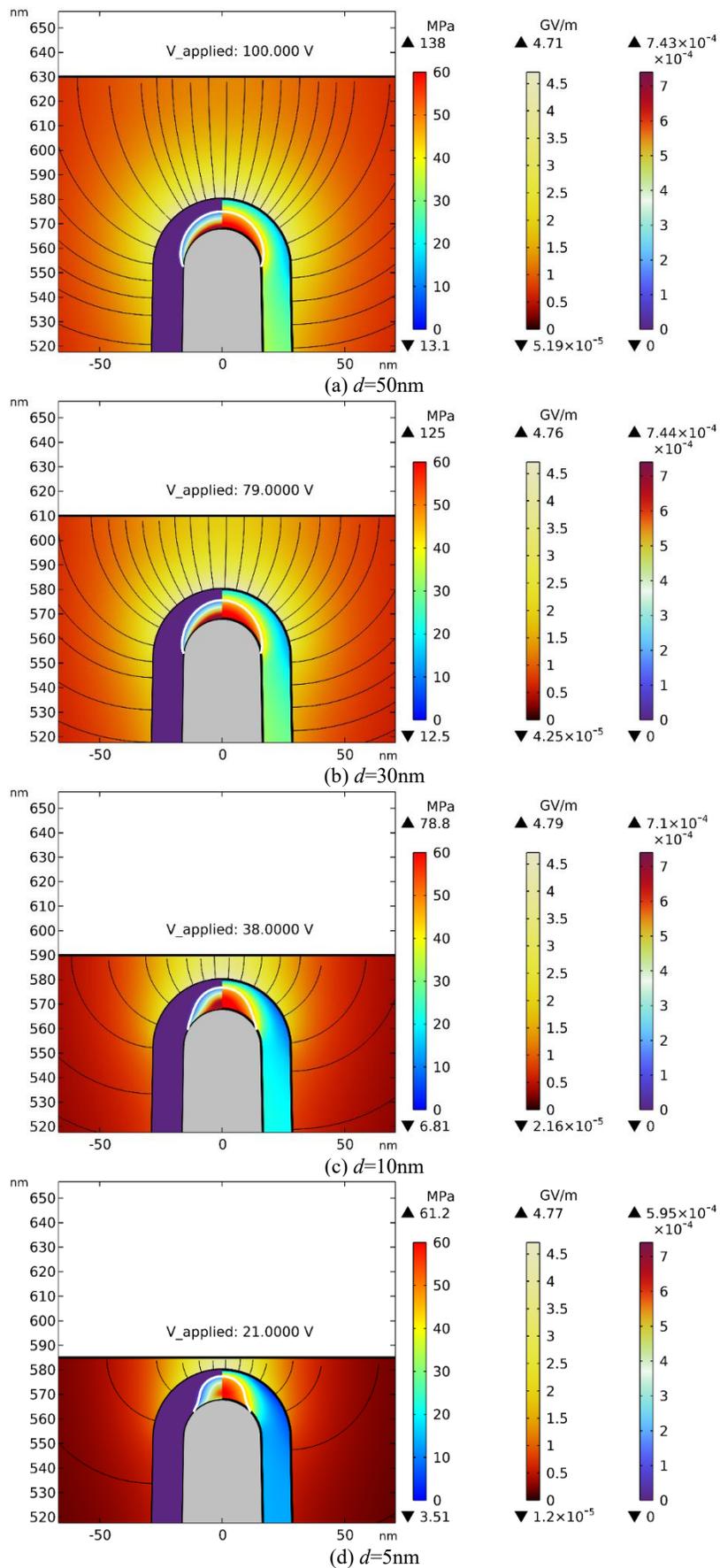

Figure S8. The simulation results of plastic deformation for various gap distances.